\documentclass[aps,pre,twocolumn,showpacs,superscriptaddress]{revtex4}
\usepackage{graphicx}

\begin{document}

\title{Rotational and Translational Phonon Modes in Glasses Composed of Ellipsoidal Particles}

\author{Peter J. Yunker}
\affiliation{Department of Physics and Astronomy, University of Pennsylvania, Philadelphia PA 19104, USA}
\author{Ke Chen}
\affiliation{Department of Physics and Astronomy, University of Pennsylvania, Philadelphia PA 19104, USA}
\author{Zexin Zhang}
\affiliation{Department of Physics and Astronomy, University of Pennsylvania, Philadelphia PA 19104, USA}
\affiliation{Complex Assemblies of Soft Matter, CNRS-Rhodia-UPenn UMI 3254}
\affiliation{Center for Soft Condensed Matter Physics and Interdisciplinary Research, Soochow University, Suzhou 215006, China}
\author{Wouter G. Ellenbroek}
\affiliation{Department of Physics and Astronomy, University of Pennsylvania, Philadelphia PA 19104, USA}
\author{Andrea J. Liu}
\affiliation{Department of Physics and Astronomy, University of Pennsylvania, Philadelphia PA 19104, USA}
\author{A. G. Yodh}
\affiliation{Department of Physics and Astronomy, University of Pennsylvania, Philadelphia PA 19104, USA}

\date{\today}
\begin{abstract}
The effects of particle shape on the vibrational properties of colloidal glasses are studied experimentally.  `Ellipsoidal glasses' are created by stretching polystyrene spheres to different aspect ratios and then suspending the resulting ellipsoidal particles in water at high packing fraction.  By measuring displacement correlations between particles, we extract vibrational properties of the corresponding ``shadow'' ellipsoidal glass with the same geometric configuration and interactions as the `source' suspension but without damping.  Low frequency modes in glasses composed of ellipsoidal particles with major/minor axis aspect ratios $\sim$1.1 are observed to have predominantly rotational character. By contrast, low frequency modes in glasses of ellipsoidal particles with larger aspect ratios ($\sim$3.0) exhibit a mix of rotational and translational character. All glass samples were characterized by a distribution of particles with different aspect ratios. Interestingly, even within the same sample it was found that small-aspect-ratio particles participate relatively more in rotational modes, while large-aspect-ratio particles tend to participate relatively more in translational modes.
\end{abstract}

\maketitle
Although the ``glass transition'' occurs in a broad array of disordered systems, including molecular [1], polymer [2], granular [3], and colloidal glasses [4], much of the physics of granular and colloidal glasses has been derived from investigation of ensembles of its simplest realization: spheres. The constituent particles of many relevant glasses, however, are anisotropic in shape or have orientation-dependent interactions. Such anisotropies are believed to affect many properties of glasses [5-10]. Thus, exploration of glasses composed of anisotropic particles holds potential to uncover new consequences for both the physics of glasses and materials applications [11].\\
\indent In glasses composed of frictionless spherical constituents, rotations of the spheres do not cost energy. Rotational modes therefore correspond to zero-frequency phonon excitations in the harmonic approximation. For anisotropic constituents, however, rotations are more energetically costly and can couple to translations. Glass vibrational properties, including the phonon density of states, are therefore expected to depend on the major/minor-axis aspect ratio of constituent particles. Simulations of disordered systems with aspect ratios marginally greater than $1.0$, for example, find low energy rotational modes that are largely decoupled from translational modes [12,13]; apparently, when particles rotate in such systems, neighboring particles rotate too, but their positions remain essentially unperturbed.\\
\indent Here we experimentally study glasses composed of ellipsoidal particles with aspect ratios, $\alpha$, ranging from $1.0-3.0$.  To this end we extend the displacement correlation matrix techniques employed in recent papers [14-17] to include rotations, and we employ video microscopy to derive the phonon density of states of corresponding ``shadow'' ellipsoidal glasses with the same geometric configuration and interactions as the experimental colloidal system but absent damping [15].  The spectra and character of vibrational modes in these disordered media was observed to depend strongly on particle aspect ratio and particle aspect ratio \textit{distribution}.  For glasses composed of particles with small median aspect ratios of $\sim$$1.1$, the lower-frequency modes are almost completely rotational in character, while higher-frequency ones are translational. In glasses of particles with larger aspect ratios ($\sim$$3.0$), significant mixing of rotations with translations is observed. In contrast to numerical findings for zero-temperature systems [12,13], we observe that the very lowest frequency modes for both systems have a mixed rotational/translational character, independent of aspect ratio. Additionally, even within the same sample, it was found that small-aspect-ratio particles tend to participate relatively more in rotational modes, while large-aspect-ratio particles tend to participate relatively more in translational modes. Evidently, the distribution of particle aspect ratios significantly affects phonon modes of glasses.\\

\begin{figure*}[!t]
\includegraphics[width=\textwidth]{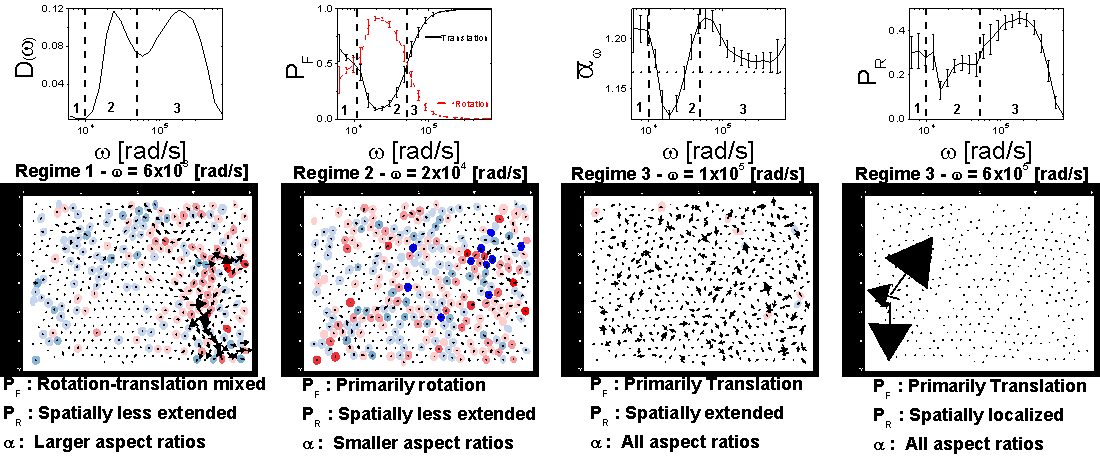}
\caption{ a. Distribution of particle aspect ratio, N($\alpha$), in samples with peak aspect ratio $\alpha_{Peak} =\ 1.1$. Inset: Experimental snapshot of part of sample. b. Vibrational density of states. Dashed lines separate three distinct regimes corresponding to modes in the vector plots displayed in f-i. c. Translational (solid black line) and rotational (dashed red line) contributions to participation fraction (P$_{F}$) plotted versus frequency $\omega$. d. Participation-fraction-averaged aspect ratio, $\bar\alpha_{\omega}$, plotted versus frequency $\omega$. e. Participation ratio (P$_{R}$) plotted versus frequency $\omega$.  f-i. Displacement vector plots of eigenmodes from lowest frequency (f) to highest (i). The size of each arrow is proportional to the \emph{translational} displacement of the particle at that position. The color intensity of each particle is proportional to the \emph{rotational} displacement of the particle at that position (with red clockwise, blue counter-clockwise, faint color is small rotation). Aspect ratio and frequency are specified in each plot.
\label{exp_info}}
\end{figure*}

\indent The experiments employ micron-sized polystyrene particles (Invitrogen) stretched to different aspect ratios [18-20].  Briefly, $3\mu m$ diameter polystyrene particles are suspended in a polyvinyl alcohol (PVA) gel and are then heated above the polystyrene melting point ($\sim$$120^{\circ}\ C$) but below the PVA melting point ($\sim$$180^{\circ}\ C$).  In the process, the polystyrene melts, but the PVA gel only softens.  The PVA gel is then placed in a vise and stretched. The spherical cavities that contain liquid polystyrene are stretched into ellipsoidal cavities.  When the PVA gel cools, the polystyrene solidifies in the distorted cavities, and becomes frozen into an ellipsoidal shape.  The hardened gel dissolves in water, and the PVA is easily removed via centrifugation. Each iteration creates $\sim$$10^{9}$ ellipsoidal particles in $\sim$$50\mu L$ suspensions. Experiments are performed on samples stretched to $110\%$ and $300\%$ of their original size (snapshots of experimental particles are shown in Fig.\ 1a, 2a insets).  The stretching scheme produces a distribution of aspect ratios with standard deviation $\sim$$18\%$. This distribution is most important for suspensions that are only slightly distorted from their initial spherical shape and therefore have greater propensity to crystallize. The distribution of aspect ratio, N$(\alpha)$, for suspensions with more spherical particles (Fig.\ 1 a) is peaked at $\alpha_{Peak} = 1.1$, with mean aspect ratio $\bar{\alpha}=1.2$, but it also has a long tail extending to $\alpha\sim2.0$.  A similar plot is shown in Fig.\ 2a for samples with $\alpha_{Peak} = 3.0$ and $\bar{\alpha}=3.3$.\\
\indent Particles are confined between glass plates to quasi-two-dimensional chambers. From separate brightness calibration studies, we estimate the chambers to be no more than $5\%$ larger than the minor axis particle length [15]. In all samples, dynamics are arrested and the spatial correlation functions of bond-orientational order decay exponentially [21], with an average bond-orientational order parameter of 0.3 (0.03) for $\alpha_{Peak}=1.1\ (3.0)$.\\
\indent Previous works have noted that the packing fraction at the jamming transition varies with particle shape [6]. In order to characterize how close our samples are to the jamming transition, we slowly evaporated water from the sample chamber. Complete evaporation should pack particles at the jamming transition for hard particles.  We verified this claim for bidisperse mixtures of spheres of size ratio 1.4, where we find $\phi _{A,MAX} = 0.84(1)$, as expected. For ellipsoids with $\alpha_{Peak}=1.1$, $\phi _{A,MAX} = 0.87(1)$, consistent with [6,9,22]; the sample employed in this paper has $\phi _{A} = 0.86(1)$. For ellipsoids with $\alpha_{Peak}=3.0$, $\phi _{A,MAX} = 0.84(1)$, again consistent with [6,9,22], while the sample employed in this paper has $\phi _{A} = 0.83(1)$. Thus both samples are near, but below, the jamming transition, with $\phi_{A,MAX}-\phi_A \approx 0.01$.

\begin{figure*}[!t]
\includegraphics[width=\textwidth]{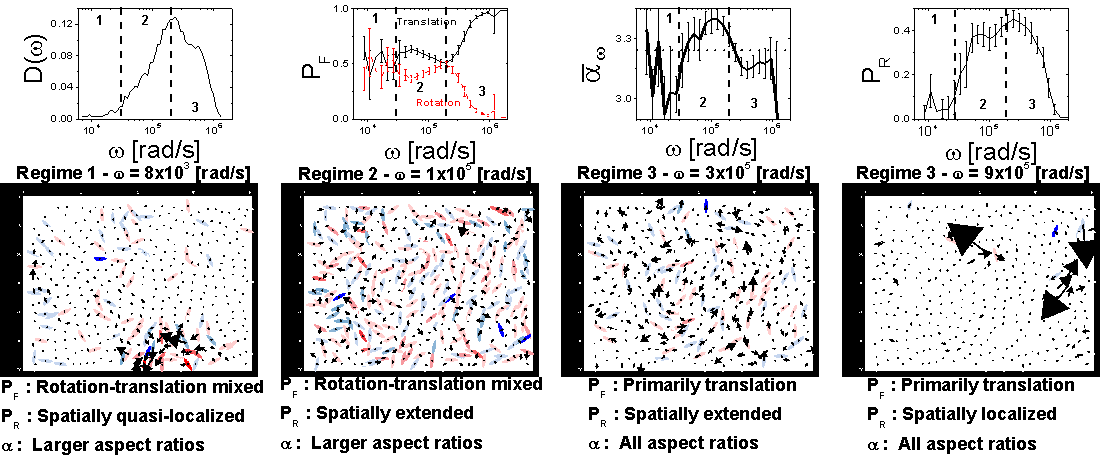}
\caption{ a. Distribution of aspect ratio, N($\alpha$), in samples with peak aspect ratio $\alpha_{Peak} =\ 3.0$. Inset: Experimental snapshot of part of sample. b. Vibrational density of states. Dashed lines separate three distinct regimes corresponding to modes in the vector plots displayed in f-i. c. Translational (solid black line) and rotational (dashed red line) contributions to participation fraction (P$_{F}$) plotted versus frequency $\omega$. d. Participation-fraction-averaged aspect ratio, $\bar\alpha_{\omega}$, plotted versus frequency $\omega$. e. Participation ratio (P$_{R}$) plotted versus frequency $\omega$. f-i. Displacement vector plots of eigenmodes from lowest frequency (f) to highest (i). The size of each arrow is proportional to the \emph{translational} displacement of the particle at that position. The color intensity of each particle is proportional to the \emph{rotational} displacement of the particle at that position (with red clockwise, blue counter-clockwise, faint color is small rotation). Aspect ratio and frequency are specified in each plot.
\label{exp_info}}
\end{figure*}

We extract vibrational properties by measuring displacement correlations.  Specifically, we define $u(t)$ as the $3N$-component vector of the displacements of all particles from their average positions ($\bar{x},\bar{y}$)and orientations ($\bar{\theta}$)($u(t)=(x(t)-\bar{x}, y(t)-\bar{y}, \theta(t)-\bar{\theta})$), and extract the time-averaged displacement correlation matrix (covariance matrix), $C_{ij} = \langle u_{i} u_{j} \rangle _{t}$ where $i, j = 1,..., 3$$N_{tot}$  run over particles, positional and angular coordinates, and the average runs over time.  In the harmonic approximation,  the correlation matrix is directly related to the sample's stiffness matrix, defined as the matrix of second derivatives of the effective pair interaction potential with respect to particle position and angle displacements.  In particular, $(C^{-1})_{ij}k_{B}T  = K_{ij}$ where $K_{ij}$ is the stiffness matrix.  Experiments that measure $C$ therefore permit us to construct and derive properties of a ``shadow'' ellipsoidal glass system that has the same static properties as our colloidal system (e.g., same correlation matrix, same stiffness matrix, but no damping) [15].  Following [23], we expect undamped hard particles that repel entropically, near but below the jamming transition, to give rise to solidlike vibrational behavior on time scales long compared to the collision time but short compared to the time between particle rearrangement events [14,17].  The stiffness matrix arising from entropic repulsions is directly related to the dynamical matrix characterizing vibrations, $D_{ij } = \frac{K_{ij}}{m_{ij}}$, where $m_{ij} = \sqrt{m_{i} m_{j}}$ and $m_{i}$ is an appropriate measure of inertia.  For translational degrees of freedom $m_{i} = m$, where $m$ is the particle mass.  For rotational degrees of freedom, $m_{i}=I_{i}$ represents the particle moment of inertia with respect to axes centered about each particle's center of mass and pointing in the $z$-direction; $I_{i} = m (a_{i}^{2} + b_{i}^{2})/2$, where $a_{i}$ and $b_{i}$ are the major and minor radii of the $i$th ellipsoid.  The eigenvectors of the dynamical matrix correspond to amplitudes associated with the various phonon modes, and the eigenvalues correspond to the frequencies/energies of the corresponding modes. Data were collected over $10,000$ seconds so that the number of degrees of freedom, $3N \approx 2000$, is small compared to the number of time frames ($\sim$$8000$) [15].  Additionally, we find $K_{ij}$ is far above the noise only for adjacent particles, as expected.

The vibrational density of states, $D( \omega)$, is plotted in Fig.\ 1b for the system with $\alpha_{Peak}\ =\ 1.1$.  $D( \omega )$ exhibits two distinct peaks.  Zero-temperature simulations find that these peaks split completely for $\alpha$ sufficiently close to 1 and for sufficiently small systems close enough to the jamming transition [12,13].  For ellipsoids with $\alpha_{Peak} = 3.0$ (Fig.\ 2b), on the other hand, $D( \omega )$ has a single peak, consistent with numerical predictions [12,13]. Thus, the vibrational spectrum of ellipsoids with small anisotropy is significantly different from those of spheres and those of ellipsoids with higher aspect ratio.

To quantitatively explore the modes, we calculated several different quantities. We will first introduce all of these quantities, and then discuss them all at once. First, to quantify the translational and rotational contributions to each mode, we sum the participation fractions, P$_{F}$, of translational and rotational vibrations over all particles, for each mode.  The eigenvectors of each mode are normalized such that $\sum _{m,n} e_{\omega}(m,n)^{2} = 1$, where $m$ runs over all particles and $n$ runs over all coordinates.  The participation \textit{fraction} for particle $m$, component $n$, in mode with frequency $\omega$ is then $P_{F}(\omega) = e_{\omega}(m,n)^{2}$.  Thus, the \textit{translational} participation fraction in a mode with frequency $\omega$ is $P_{F,XY}(\omega) = \sum_{m=1..N,n=X,Y}e_{\omega}(m,n)^{2}$ and the \textit{rotational} participation fraction is $P_{F,\theta}(\omega) = 1-P_{F,XY}(\omega)=\sum_{m=1..N}e_{\omega}(m,\theta)^{2}$.  Translational and rotational participation fractions are plotted in Fig.\ 1c and Fig.\ 2c.

To investigate effects of polydispersity in the aspect ratio, $\alpha$, we measure the eigenvector-weighted ellipsoid aspect ratio as a function of mode frequency. Specifically, we compute $\bar\alpha_{\omega}\ =\ \sum_{m,n}\ \alpha_{m}e_{\omega}(m,n)^{2}$, where $\alpha_{m}$ is the measured aspect ratio of particle $m$.  $\bar\alpha_{\omega}$ is thus a measure of the average particle aspect ratio for the particles participating in mode $\omega$ (Figs.\ 1d and 2d).

Finally, to assess the degree of mode localization, we quantify the spatial extent of individual modes by computing the participation \textit{ratio}, $P_{R}(\omega) = (\sum _{m,n} e _{\omega}(m,n) ^{2})^{2} / (N_{tot} \sum _{m,n} e _{\omega}(m,n) ^{4})$ (Fig.\ 1e and Fig.\ 2e).  The participation ratio provides an indication of mode localization in space.  If a mode is localized, a small number of terms will dominate, making $\sum _{m,n} e _{\omega}(m,n) ^{4}$ and $(\sum _{m,n} e _{\omega}(m,n) ^{2})^{2}$ similar in size so $P_{R}( \omega) \approx 1 / N$.

Representative modes are shown in Figs.\ 1f-i, 2f-i for samples with $\alpha_{Peak}=1.1$ and $3.0$, respectively. Modes from all samples fall qualitatively into three regimes. For $\alpha_{Peak} = 1.1$, three distinct regimes exhibiting different behavior are labeled in Fig.\ 1b-e.  For mode frequencies higher than $\omega \approx 54000$ rad/s, i.e., frequencies above the ``dip" separating the two peaks in the density of states (Fig.\ 1b), the modes (regime 3) are translational in character. Interestingly, the lowest frequency modes in regime 3 are spatially extended (Fig.\ 1h), while the highest frequency modes are spatially localized (Fig.\ 1i), similar to the modes in glasses composed of spheres.  Modes just above $\omega \approx 54000$ rad/s are enriched in longer ellipsoids and have a mixed translational/rotational character.  Modes in regime 2, extending from $1300 \lesssim \omega \lesssim 54000$ rad/s, are strongly rotational in character and are concentrated on small aspect-ratio particles (Fig.\ 1g).   In regime 1, below $\omega \approx 1300$ rad/s, modes again have a mixed rotational/translational character and are concentrated on longer particles (Fig.\ 1f). Regime 1 was not observed in numerical simulations of monodisperse ellipsoid packings at zero temperature [12,13].  We also find that the mean value of elements of the stiffness matrix connecting particles to their neighbors decreases as aspect ratio increases [21]; this observation suggests that longer ellipsoids are more weakly coupled to their neighbors and are relatively more likely to be excited at low frequency.

\begin{figure}
\scalebox{1.0}{\includegraphics{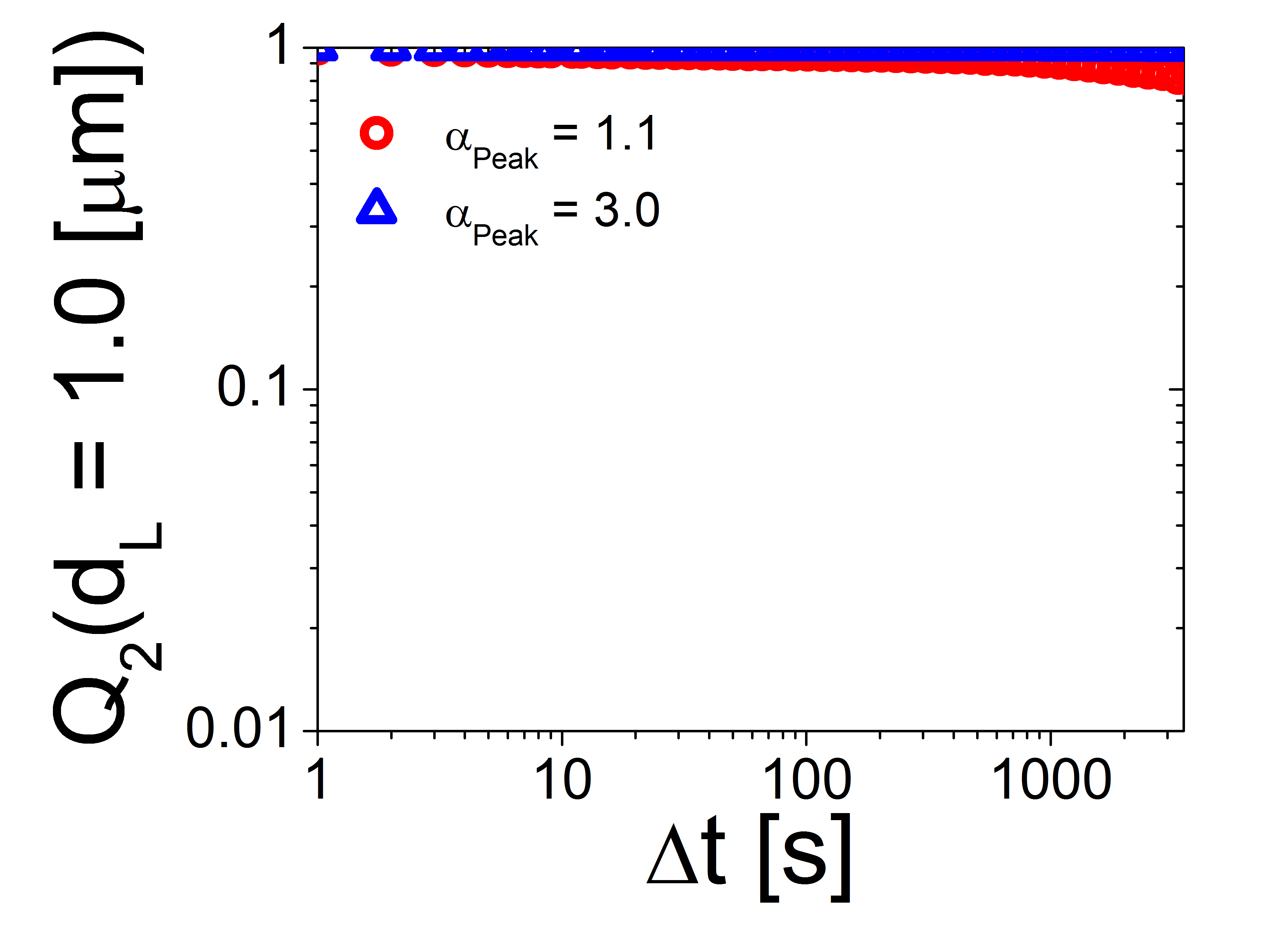}}
\caption{ The two-point-correlation function, Q$_{2}$, which probes self overlap, is plotted versus delay time for ellipsoidal glasses with different aspect ratios. Dynamic arrest is apparent.
\label{exp_info}}
\end{figure}

Figs.\ 2b-e show that for $\alpha_{Peak} = 3.0$, high frequency modes above $\omega \approx 3 \times 10^5$ rad/s in regime 3 are translational in character with a nearly average mode-averaged aspect ratio, resembling those of spheres. These translational modes cross from extended (Fig.\ 2h) to localized (Fig.\ 2i) at the upper end of the spectrum.  Modes with $20000 \lesssim \omega \lesssim 3 \times 10^5$ rad/s in regime 2 are extended with a mixed rotational/translational character and are slightly enriched with longer ellipsoids at higher frequencies and shorter ellipsoids at somewhat lower frequencies (Fig.\ 2g).  In regime 1, $\omega \lesssim 2 \times 10^{4}$ rad/s, modes are again slightly enriched in longer-aspect ratio particles and are quasilocalized with mixed translational/rotational character (Fig.\ 2f).

Comparing the two systems, the behavior of modes at high frequencies (regime 3) and at the lowest frequencies (regime 1) are qualitatively very similar. The largest qualitative differences between large and small aspect ratio systems occurs in regime 2, where modes have primarily rotational character for systems with $\alpha_{Peak} = 1.1$ and modes have mixed translational/rotational character for systems with $\alpha_{Peak} = 3.0$.

To summarize, experiments suggest the nature of low frequency modes in glasses depends strongly on constituent particle aspect ratio.  Rotational modes tend to occur at lower frequencies than translational vibrations, and, for glasses with aspect ratios $\sim$$1.1$, a frequency regime exists wherein the spectrum is strongly rotational in character, consistent with numerical results [12,13]. Additionally, even within each sample, particles with small aspect ratios tend to participate more in rotational modes, while particles with larger ones tend to participate more in translational modes. We also find low frequency modes enhanced in longer particles with mixed rotational/translational character that were not present in simulations. The \textit{distribution} of particle aspect ratio, N($\alpha$), is thus an important physical factor affecting phonon modes of ellipsoidal glasses. Recent work suggests that low-participation-ratio, low-frequency modes appear to correlate with regions prone to rearrangement or plastic deformation [24].  Thus, the existence of additional low frequency modes concentrated around particles with certain aspect ratios may have important consequences for the mechanical response of glasses.

\begin{figure}
\scalebox{1.0}{\includegraphics{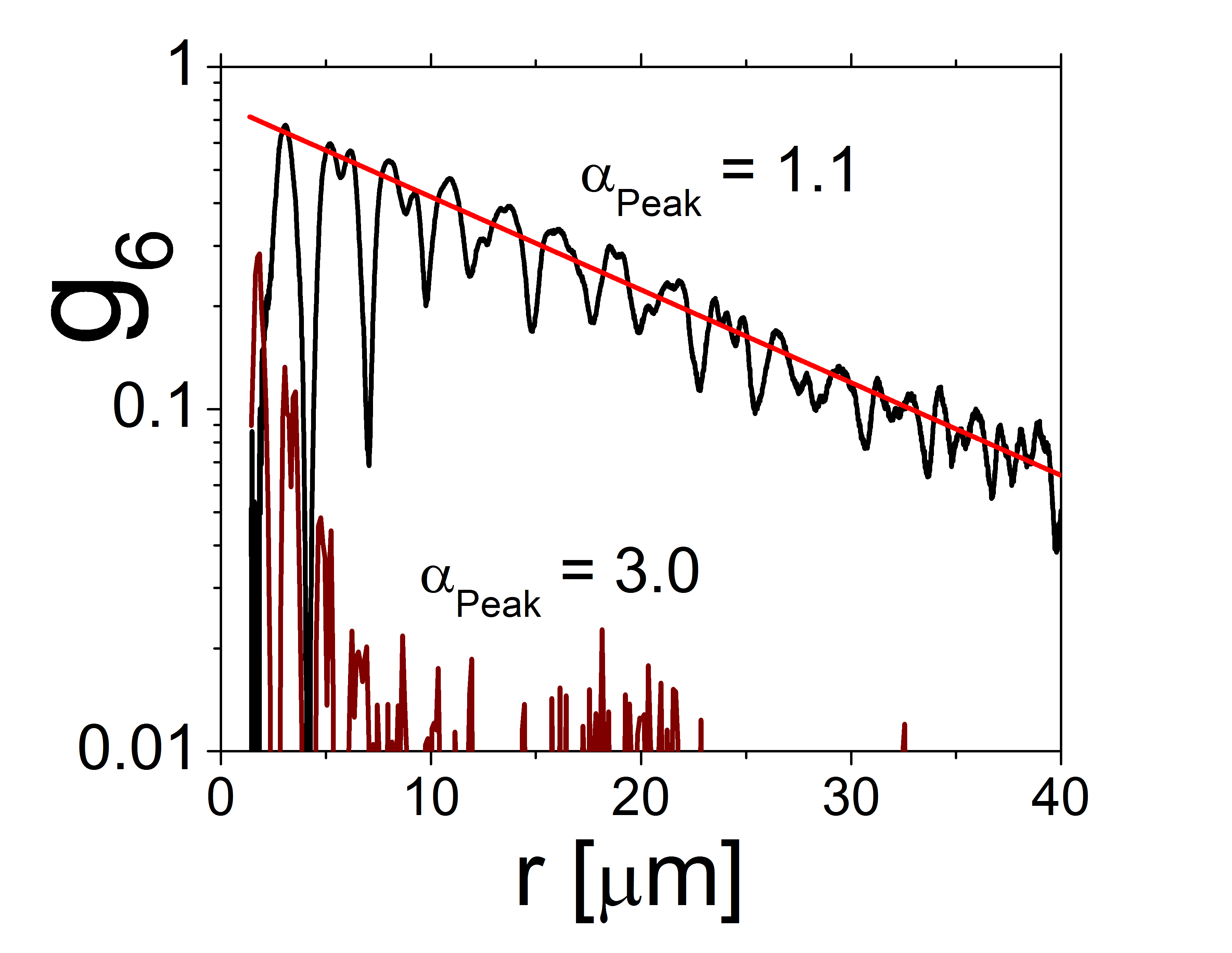}}
\caption{ Bond orientational order spatial correlation functions, g$_{6}$(r), for ellipsoidal glasses with different aspect ratios.
\label{exp_info}}
\end{figure}

\begin{acknowledgments}
We thank Kevin B. Aptowicz, Dan Chen, Piotr Habdas, and Matthew Lohr for helpful discussions, and we gratefully acknowledge financial support from the National Science Foundation through DMR-0804881, the PENN MRSEC DMR-0520020, and from NASA NNX08AO0G.
\end{acknowledgments}

\section{Appendix A: Glassy Dynamics}
As a first step towards elucidation of glass dynamics in these systems, we compute the two-time self-overlap correlation function:
$Q_{2}(d_{L},\Delta t)=\frac{1}{N_{tot}}\sum_{i=1}^{N_{tot}}exp(-\frac{\Delta r_{i}(\Delta t)^{2}}{2d_{L}^{2}})$ (Fig.\ 3) [25].  Here $d_{L}$ is a pre-selected length scale to be probed, $N_{tot}$ is the total number of particles, and $\Delta r_{i}(\Delta t)$ is the distance particle $i$ moves in time $\Delta t$. If a particle moves a distance smaller than $d_{L}$, $Q_{2}$ will remain approximately unity; if a particle moves a distance greater than $d_{L}$, $Q_{2}$ will fall to zero.  Notice that for glasses of each aspect ratio, $Q_{2} (d_{L} = 1.0$ $\mu m)$ decays very little over the experimental timescale, thereby indicating that glass dynamics are arrested at length scales of order the particle-size.

\begin{figure}
\scalebox{1.0}{\includegraphics{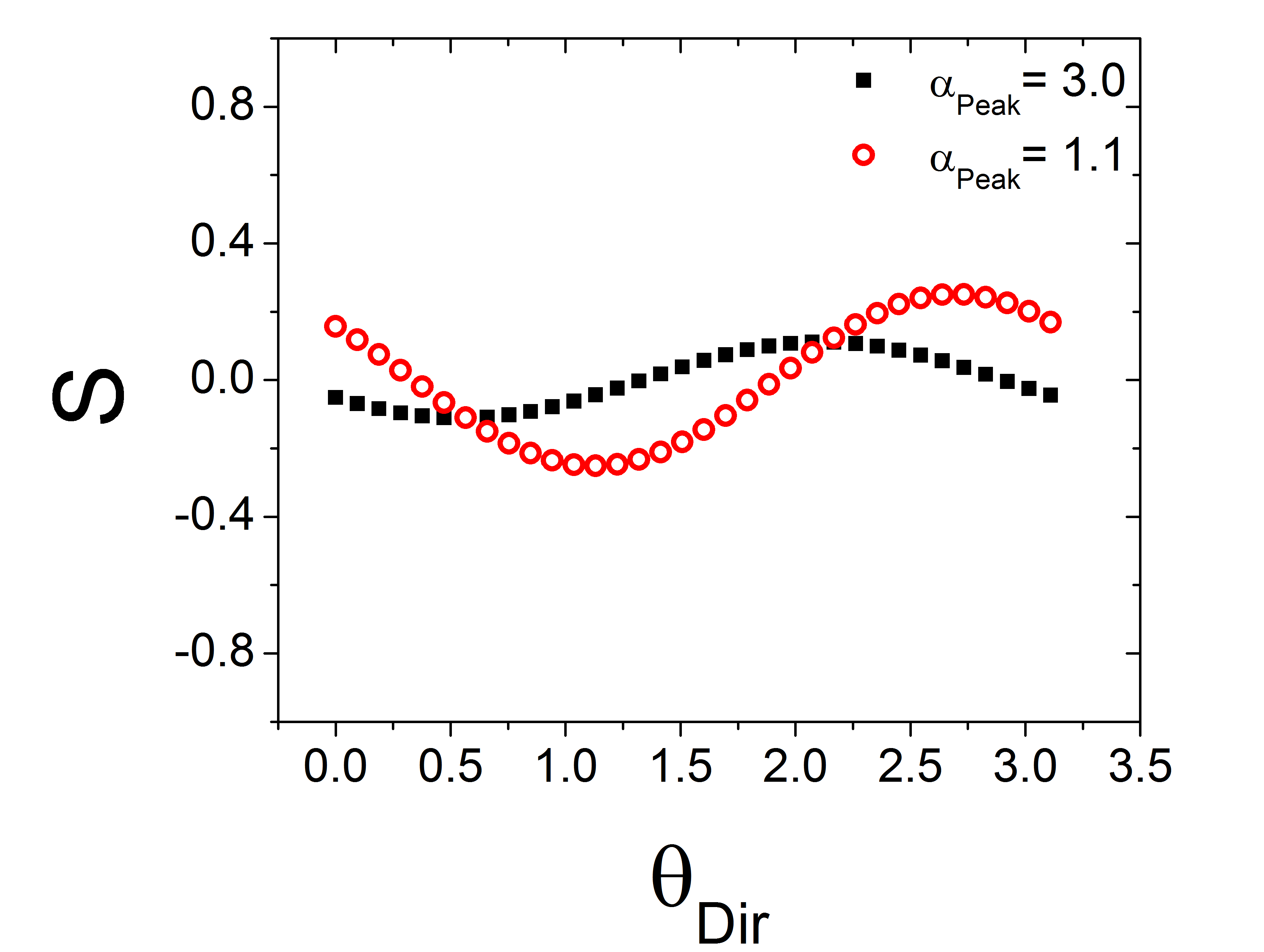}}
\caption{ The average nematic order parameter, S, as a function of the director angle for samples with $\alpha_{Peak} = 3.0$ (closed squares) and $\alpha_{Peak} = 1.1$ (open circles).
\label{exp_info}}
\end{figure}

\section{Appendix B: Bond-Orientational Order}
To demonstrate the absence of long-range orientational order in these systems, the bond-orientational order parameter, $\psi_{6}=\frac{1}{N_{tot}CN}\sum_{j=1}^{N_{tot}}|\sum_{k=1}^{CN} e^{i6\theta_{jk}}|$ and its spatial correlation function $g_{6}$$(r$=$|\textbf{r$_{i}$-r$_{j}$}|)=\langle\psi_{6i}^{*}(r_{i})\psi_{6j}(r_{j})\rangle$ are calculated (Fig.\ 4). Here $\theta$$_{jk}$ is the angle between the x-axis and the $j-k$ bond between particles $j$ and $k$, CN is the coordination number of particle $j$, and $r_{i}$ and $r_{j}$ are the positions of particles $i$ and $j$.  $g_{6}$ decays faster in samples with $\alpha_{Peak}=3.0$ than it does in samples with $\alpha_{Peak}=1.1$. However, $g_{6}$ decays exponentially in each sample (see exponential fit line in Fig.\ 4), a signature of structural disorder characteristic of glasses (e.g. [26]).

\section{Appendix C: Nematic Order}
To demonstrate the absence of long-range nematic order in these systems, the nematic order parameter, $S=\sum_{j=1}^{N_{tot}}2*cos(\theta_{j}-\theta_{Dir})^{2}-1$, where $\theta_{j}$ is the orientation of particle i and $\theta_{Dir}$ the orientation of the nematic director, and angle brackets represent ensemble averaging, is largely absent (Fig. 5).  For an isotropic distribution of orientations, S = 0, and for perfectly aligned particles S = 1.  The mean value of S in our high aspect ratio samples ($\alpha_{Peak} = 3.0$) is 0.05, and the maximum value of S is 0.11. The mean value of S in samples with $\alpha_{Peak} = 1.1$ is 0.00, and the maximum value of S is 0.25.

\section{Appendix D: Low Frequency Modes with Mixed Roational/Translational Character}
Low frequency modes for samples with $\alpha_{Peak}=1.1$ have mixed rotational/translational character. These modes were not seen in zero-temperature simulations in which all particles have identical aspect ratios [12,13]. These `mixed' modes typically involve larger aspect ratio particles. To understand why these modes appear at low frequencies, we calculated the average spring constant connecting a particle's rotation to its nearest neighbors $K_{iNN}=<K_{ij}/m_{ij}>_{NN}$, where $<>_{NN}$ indicates an average over nearest neighbors pairings, $i$ runs over all theta components and $j$ runs over all components. We then plotted $K_{iNN}$ as a function of aspect ratio (Fig.\ 6). $K_{iNN}$ decreases as $\alpha$ increases, indicating that the average spring constraining rotation decreases as $\alpha$ increases. Smaller spring constants $K_{iNN}$ lead to vibrations at smaller frequencies. Thus, particles with longer aspect ratios tend to vibrate at lower frequencies.

\begin{figure}
\scalebox{1.0}{\includegraphics{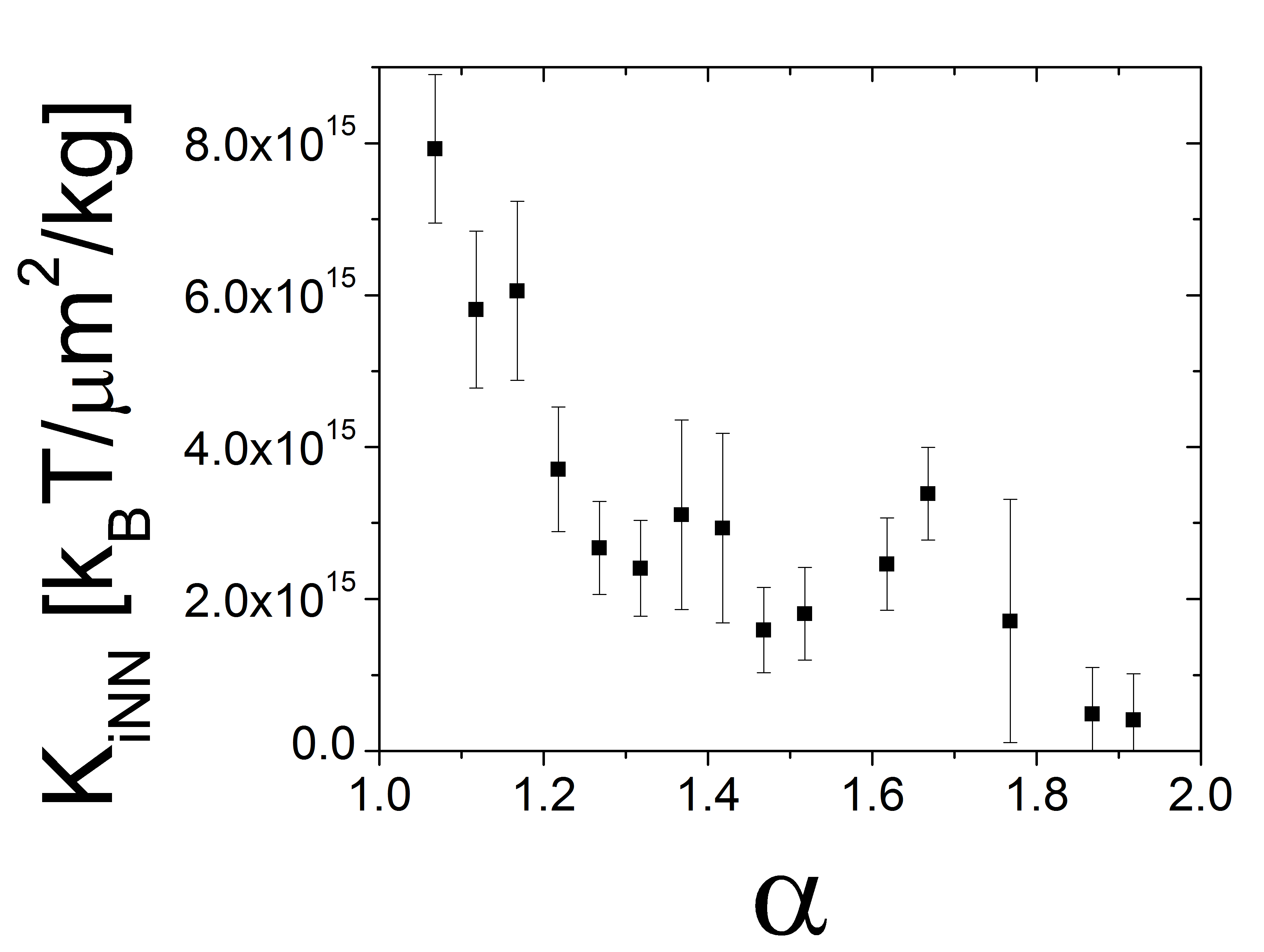}}
\caption{ The average spring constant, $K_{iNN}$, connecting nearest neighbors as a function of aspect ratio, $\alpha$, for samples with $\alpha_{Peak}=1.1$. Error bars represent standard error.
\label{exp_info}}
\end{figure}

\noindent [1] C. A. Angell, Science \textbf{267}, 1924 (1995).\newline
[2] J. Hutchinson, Prog. Polym. Sci. \textbf{20}, 703 (1995).\newline
[3] A. R. Abate and D. J. Durian, Phys. Rev. Lett. \textbf{101}, 245701 (2008).\newline
[4] E. Weeks, et al., Science \textbf{287}, 627 (2000).\newline
[5] A. Jaoshvili, et al., Phys. Rev. Lett. \textbf{104}, 185501 (2010).\newline
[6] A. Donev, et al., Science \textbf{303}, 990 (2004).\newline
[7] W. Man, et al., Phys. Rev. Lett. \textbf{94}, 198001 (2005).\newline
[8] M. J. Solomon and P. T. Spicer, Soft Matter \textbf{6}, 1391 (2010).\newline
[9] A. Donev, et al., Phys. Rev. E \textbf{75}, 051304 (2007).\newline
[10] R. C. Kramb, et al., Phys. Rev. Lett. \textbf{105}, 055702 (2010).\newline
[11] D. A. Weitz, Science \textbf{303}, 968 (2004).\newline
[12] M. Mailman, et al., Phys. Rev. Lett. \textbf{102}, 255501 (2009).\newline
[13] Z. Zeravcic, et al., Euro. Phys. Lett. \textbf{87}, 26001 (2009).\newline
[14] A. Ghosh, et al., Soft Matter \textbf{6}, 3082 (2010).\newline
[15] K. Chen, et al., Phys. Rev. Lett. \textbf{105}, 025501 (2010).\newline
[16] D. Kaya, et al., Science \textbf{329}, 656 (2010).\newline
[17] A. Ghosh, et al., Phys. Rev. Lett. \textbf{104}, 248305 (2010).\newline
[18] B. Felder, Helvetica Chimica Acta \textbf{49}, 440 (1966).\newline
[19] C. C. Ho, et al., Colloid and Polymer Science \textbf{271}, 469 (1993).\newline
[20] J. A. Champion, Y. K. Katare, and S. Mitragotri, Proc. Nat. Acad. Sci. USA \textbf{104}, 11901 (2007).\newline
[21] For more information see online supplemental material.\newline
[22] C. F. Schreck, N. Xu, and C. S. O'Hern, Soft Matter \textbf{6}, 2960 (2010).\newline
[23] C. Brito and M. Wyart, Eur. Phys. Lett. \textbf{76}, 149 (2006).\newline
[24] A. Widmer-Cooper, et al., Nat. Phys. \textbf{4}, 711 (2008).\newline
[25] O. Dauchot, G. Marty, and G. Biroli, Phys. Rev. Lett. \textbf{95}, 265701 (2005).\newline
[26] H. Tanaka, T. Kawasaki, H. Shintani, and K. Watanabe, Nat. Mater. \textbf{9}, 324 (2010).\newline

\end{document}